\documentclass[twocolumn,showkeys,showpacs,preprintnumbers,amsmath,amssymb,prl,superscriptaddress,nofootinbib]{revtex4}
\usepackage{amsmath}
\usepackage{graphicx}
\usepackage{dcolumn}% Align table columns on decimal point \usepackage{bm}%
\usepackage{bm} %bold math

\newcommand{\degree}{{$^\circ$}}
\begin{document}
%\preprint{APS/}
%\title{\boldmath New Yb-heavy fermion compound with new 1-2-4 structure type in the vicinity of quantum critical point}
\title{\boldmath Quantum Criticality Based on Large Ising Spins: YbCo$_2$Ge$_4$ with New 1-2-4 Structure Type}

\author{K.~Kitagawa}
\email{kitag@kochi-u.ac.jp}
\affiliation{Graduate School of Integrated Arts and Sciences, Kochi University, Kochi 780-8520, Japan}
\author{Y.~Kishimoto}
\affiliation{Graduate School of Integrated Arts and Sciences, Kochi University, Kochi 780-8520, Japan}
\author{M.~Iwatani}
\affiliation{Graduate School of Integrated Arts and Sciences, Kochi University, Kochi 780-8520, Japan}
\author{T.~Nishioka}
\affiliation{Graduate School of Integrated Arts and Sciences, Kochi University, Kochi 780-8520, Japan}
\author{M.~Matsumura}
\affiliation{Graduate School of Integrated Arts and Sciences, Kochi University, Kochi 780-8520, Japan}
\author{K.~Matsubayashi}
\affiliation{Institute for Solid State Physics, University of Tokyo, Kashiwanoha, Kashiwa, Chiba 277-8581, Japan}
\author{Y.~Uwatoko}
\affiliation{Institute for Solid State Physics, University of Tokyo, Kashiwanoha, Kashiwa, Chiba 277-8581, Japan}
\author{S.~Maki}
\affiliation{Materials Research Center for Element Strategy, Tokyo Institute of Technology, Yokohama 226-8503, Japan}
\author{J-I.~Yamaura}
\affiliation{Materials Research Center for Element Strategy, Tokyo Institute of Technology, Yokohama 226-8503, Japan}
\author{T.~Hattori}
\affiliation{Department of Physics, Graduate School of Science, Kyoto University, Kyoto 606-8502, Japan}
\author{K.~Ishida}
\affiliation{Department of Physics, Graduate School of Science, Kyoto University, Kyoto 606-8502, Japan}

\date{\today}% It is always \today, today,
\begin{abstract}
We present a new type of quantum critical material YbCo$_2$Ge$_4$, having the largest quantum-critical pseudospin size ever.
The YbCo$_2$Ge$_4$-type structure is new, forms in the orthorhombic $Cmcm$ system, and is related to the well-known ThCr$_2$Si$_2$ structure.
Heavy rare earth (Tm,Yb,Lu, or Y) members are also possible to be grown.
YbCo$_2$Ge$_4$ possesses the Ising-type ground-state doublet, 
namely the simplest ones of uniaxially up or down, $|\pm \sim 7/2\rangle$. 
It is clearly manifested through comprehensive resistivity, magnetization, specific heat, and NQR/NMR experiments.
Large pseudospin state usually tends to order in simple magnetisms, or hard to be screened by Kondo effect.
Therefore, the discovery of the quantum criticality of the fluctuating large spins opens a new door to new-material search and theoretical studies.
\end{abstract}

%Useshowkeys
%classoption if keyword %display desired 
\maketitle

Diversity of quantum states---non-Felmi-liquid (non-FL) or unconventional phases like superconductivity---emerges near the zero-point phase transition. 
Electronic degrees of freedom remain fluctuating at this specific point, which is so called a quantum critical point (QCP). 
% Zero-point state of electrons organized in a crystal is, in most situations, delocalized (Landau Fermi-liquid, FL)
 %or localized (magnetic orders). 
Nevertheless, non-FL or quantum criticality (QC) is very rare phenomenon to be observed at ambient pressure in a stoichiometric (not chemically diluted) clean crystal.
Fine tuning by pressure or doping is necessary to attain QCP, except for the lucky cases that have been found only in several heavy fermion systems.
CeNi$_2$Ge$_2$, CeCu$_2$Si$_2$, UBe$_{13}$, and $\beta-$YbAlB$_4$ shows QC at ambient pressure, and all of them are superconducting\cite{LoehneysenQPTReview,MatsumotoYbAlB4}.
Thus finding a new QC material at ambient pressure, being reported here, is very important.
Another remark is that QC behavior based on ytterbium
---Yb$^{3+}$, $4f^{13}$ has a complemental electronic configuration to Ce$^{3+}$, $4f^{1}$---
has been found to be rare and simultaneously exotic,
 in contrast to great numbers of Ce/U-based antiferromagnetic (AF) QC which are in accord with the spin fluctuation theories\cite{Hertz,Millis,SCRBook}.
For example, YbRh$_2$Si$_{2-x}$Ge$_x$, near the AF QCP, shows divergent behavior of the uniform magnetic susceptibility $\chi$ and  very large Wilson ratio $R_\text{W}$, which is a dimensionless measure of ferromagnetic (FM) enhancement in FL, reaching as high as 14\cite{GegenwartYbRh2Si2,CustersYbRh2Si2}.
Intermediate-valent Yb materials, $\alpha,\beta-$YbAlB$_4$ also exhibit $R_\text{W} > 9$\cite{MatsumotoYbAlB4}.
More compatible theories for these are, for example, the local QCP\cite{SiLocalQCPJPSJ} or valence-crossover QCP\cite{WatanabeVFQCPPRL} scenarios.
The supposed QC classes have to be confirmed with more experimental evidences in Yb QCP.

A ThCr$_2$Si$_2$-type (hereafter 122-type) structure is a very typical framework for studying strongly correlated electrons, 
including the Fe-based high-$T$ superconductors, many heavy-fermion superconductors/QC materials, and YbRh$_2$Si$_2$.
The next heaviest isostructural Yb compound, YbCo$_2$Ge$_2$\cite{KolendaYbCo2Ge2,McCallYbCo2Ge2}
 has nearly trivalent Yb and fairly enhanced mass of $\gamma \sim 200$~mJ/mol-K$^2$.
However the high-pressure study on the polycrystalline form suggests that a QCP is yet away even at 3~GPa\cite{TrovarelliYbCo2Ge2}.
In this Letter, we demonstrate that incorporating extra Ge cations to 122-type structure is very effective to realize QC with Ising-type spins.
The QC nature of YbCo$_2$Ge$_4$ with a new 124 structure is revealed by the Sommerfeld coefficient of the electronic specific heat,
 $\gamma$, diverging to zero temperature. 
Further macro/microscopic magnetic experiments found that 
underlying QC fluctuations are AF, but not of a simple commensurate one.
The universality class of QC differs from those in the other previously known Yb-based QC. 
\begin{figure}[htbp]
\centering
\includegraphics[width=0.8\linewidth]{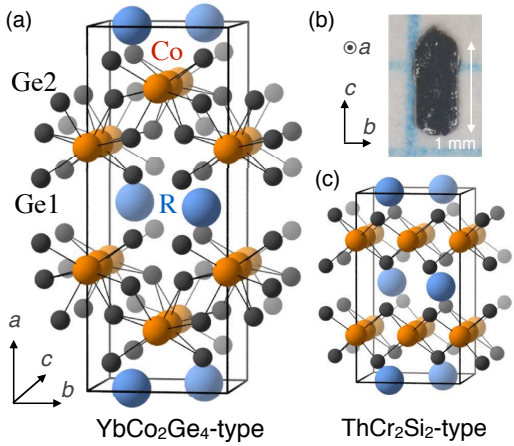}
\caption{\label{fig:struct}{(color online) Newly discovered YbCo$_2$Ge$_4$.}
(a) Crystal structure of new YbCo$_2$Ge$_4$ type and the frame for the orthorhombic unit cell.
(b) Photo of a grown YbCo$_2$Ge$_4$ single crystal.
(c) ThCr$_2$Si$_2$ structure is shown, in which the frame indicates double-sized tetragonal unit cell, for comparison.
Folding the Co-Ge conductive planes after embedding Ge1 sites into the 122 equals to the 124 structure. 
}
\end{figure}

Through search of $R$Co$_2$Ge$_4$, tin flux was the only successful method within our trials, 
Appropriate starting compositions to obtain the largest single crystals were 
$R$:Co:Ge=1:2:4 for $R$=Y,Tm,Lu, and 9:4:6 for $R$=Yb with sufficient amount of Sn.
High purity elements (99.99\% Yb or 99.9\% $R$, and 99.999\% Co, Ge, Sn) were put in BN crucibles and heated in sealed quartz tubes.
The tubes were slowly cooled down from at least 950{$^\circ$}C  
to 300{$^\circ$}C and centrifuged. 
\begin{table}[htbp]
\caption{\label{tab:scxrd}Crystallographic data of YbCo$_2$Ge$_4$ and LuCo$_2$Ge$_4$ determined by single-crystal XRD}
\begin{ruledtabular}
\begin{tabular}{ccc}
formula& YbCo$_2$Ge$_4$& LuCo$_2$Ge$_4$\\
\hline
Temperature(K)& 293& 273\\
Space group& \multicolumn{2}{c}{$Cmcm$ (No.~63)}\\
$a$ (nm)& 1.44568(1)& 1.44422(1)\\
$b$ (nm)& 0.56286(1)& 0.56290(1)\\
$c$ (nm)& 0.53992(1)& 0.53886(1)\\
$V$ (nm$^3$)& 0.43933(1)& 0.43807(1)\\
$Z$&  \multicolumn{2}{c}{4}\\
%Reflections(total)& 838& 8977\\
%Reflections& 622& 1482\\
$R$& 0.045& 0.046\\
\hline
\multicolumn{3}{c}{Atomic parameters, setting number 1}\\
\multicolumn{3}{c}{Yb/Lu $4c$ $(0, y, 0.25)$}\\
$y$& 0.7223(2)& 0.72215(6) \\
$U_{eq}$ (pm$^2$)& 72(2)& 39(1)\\
\multicolumn{3}{c}{Co $8e$ $(x,0,0)$}\\
$x$& 0.3313(1)& 0.33184(4)\\
$U_{eq}$(pm$^2$)& 79(3)& 38(1)\\
\multicolumn{3}{c}{Ge1 $8g$ $(x, y, 0.25)$}\\
$x$& 0.08857(8)& 0.08834(4)\\
$y$& 0.2137(3)& 0.2128(1)\\
$U_{eq}$ (pm$^2$)& 70(3)& 42(1)\\
\multicolumn{3}{c}{Ge2 $8g$  $(x, y, 0.25)$}\\
$x$&  0.20189(8)& 0.20214(4)\\
$y$& -0.1503(2)& -0.1503(1)\\
$U_{eq}$ (pm$^2$)& 70(3)&  50(1)\\
\end{tabular}
\end{ruledtabular}
\end{table}

Figure~\ref{fig:struct}(a) displays the new $R$Co$_2$Ge$_4$-type structure, and  
crystallographic parameters, atomic positions, displacement parameters for $R=$Yb and Lu are summarized in Table.~\ref{tab:scxrd}.
ThCr$_2$Si$_2$-type structure is also displayed in Fig.~\ref{fig:struct}(c) to emphasis the differences:
while the original 122 has straight anti-PbO-type Co-Ge conduction layers, the 124 structure has folded Co-Ge2 layers and the extra Ge1 sites.
The Ge1 sites surrounds the rare earth site, as the conduction Ge of 122 does so.
These similarities suggest a certain relationship in physical properties between $R$Co$_2$Ge$_4$ and $R$Co$_2$Ge$_2$ members.
However, an apparent difference is that the former is orthorhombic.
The $R$-site crystal electric fields (CEF), which govern the ground state (GS) of heavy fermions, reflect the local symmetry at the $R$ site.
The $4c$ site in 124 cannot have four-fold symmetry, and therefore XY-type magnetic anisotropy like YbRh$_2$Si$_2$\cite{TrovarelliYbRh2Si2}
 cannot be realized.    
Later the anisotropy of YbCo$_2$Ge$_4$ is shown to be Ising-type along the $b$ axis. 
\begin{figure}[htbp]
\centering
\includegraphics[width=0.9\linewidth]{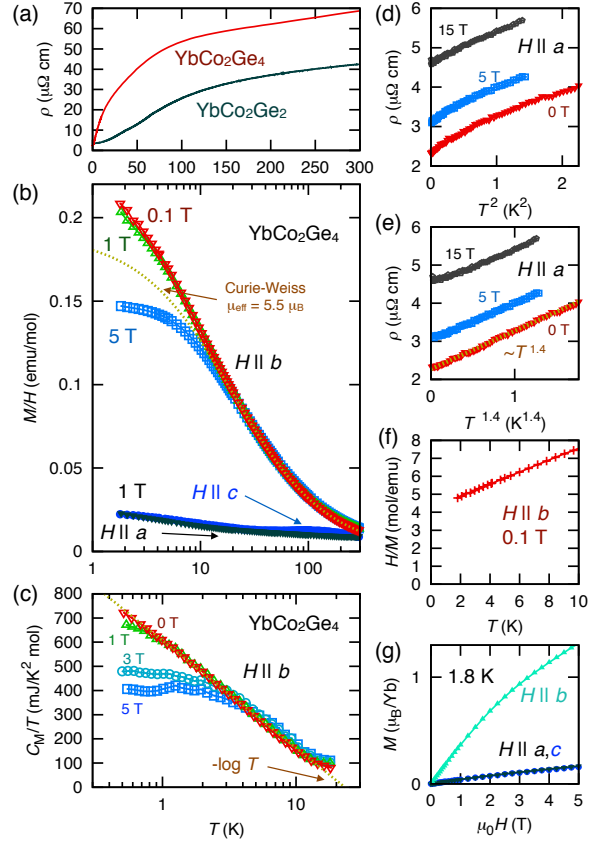}
\caption{\label{fig:bulk}{(color online) The macroscopic measurement results of YbCo$_2$Ge$_4$.}
(a) The in-plane resistivity for YbCo$_2$Ge$_4$, $\rho_c$, and YbCo$_2$Ge$_2$, $\rho_{ab}$. 
(b) Susceptibility $M/H$ for $H \parallel a,b,$ and $c$ at 0.1, 1, and 5~T.
The dotted line indicates a Curie-Weiss fit between 30~K and 200~K for the easy axis $b$. The effective moment $\mu_\text{eff}$ is 5.5~$\mu_\text{B}$ per Yb and 
a Curie-Weiss temperature $\Theta_\text{W} = -20$~K. 
(c) Magnetic specific heat divided by $T$, $C_\text{M}/T$. The linear dotted line indicates a QC behavior $\sim -\log T$. 
(d,e) $\rho_c$ at low-$T$ ($T < 1.5$~K) when the field is applied along $a$, plotted against $T^2$ [panel (d)] and $T^{1.4}$ [panel(e)].
(f) Inverse susceptibility $H/M$ plot at 0.1~T for the easy axis $b$ from low $T$ ($T < 10$~K) data of the panel (b).
Curie-Weiss law, evidenced by the straight behavior, persists down to 1.8~K.
(g) Magnetization curves at 1.8~K.
}
\end{figure}

The basic bulk properties of YbCo$_2$Ge$_4$ (from here denoted as Yb-124) and in part those for YbCo$_2$Ge$_2$ (Yb-122) and LuCo$_2$Ge$_4$ (Lu-124) are shown in Fig.~\ref{fig:bulk} and in Fig.~1 of Supplemental Material.
In Yb-124, strong electron scattering remains to the lowest temperatures ($T$), as shown in convex aperture of the resistivity $\rho$
[Fig.~\ref{fig:bulk}(a)], $T$-dependent uniform susceptibility $M/H$ (magnetization over field) [Fig.~\ref{fig:bulk}(b)], and $-\log T$-like behavior of the specific heat $C$ over $T$ [Fig.~\ref{fig:bulk}(c)].   
These make a marked contrast to Yb-122, of which $\rho$ shows a standard concave $T^2$ behavior and constant $C/T$ at low $T$ of very FL.
The residual resistivity ratio $\rho(300~\text{K})/\rho(0~\text{K})$ reaches no less than 30, 
which is good enough to examine the low-$T$ nature. 

To clarify how Yb-124 persists non-FL at low $T$ and elucidate its criticality class, expanded views of $\rho$ and $M$ are displayed in the panels (d-f).
No ordering is observed in $\rho(T)$ down to 20~mK, and the power law of $\rho$ at the lowest $T$ range (20~mK$<T<$1.5~K) is about $T^{1.4}$
[the dotted line in the panel (e)], which is non-FL.
The inverse susceptibility $H/M$ indicates that the uniform susceptibility $\chi(\bm q= 0)$ has no tendency to diverge toward 0~K,
 and then the FM mode is not QC.
Nevertheless, the high degeneracy of the GS caused by the QC is clearly observed in the $C_\text{M}/T = \gamma$, 
which is the $C/T$ difference between Yb-124 and Lu-124.
The diverging $\gamma \propto log(T^*/T)$ behavior is shown as the dashed line in Fig.~\ref{fig:bulk}(c).
A single ion Kondo or characteristic spin fluctuation scale,  $T^*$, of 23~K, is inferred from the fit, 
and 77\% of the entropy of the GS doublet, $R \log 2$,  is released at $T^*$. 
This $T^*$ is comparable to that of YbRh$_2$Si$_2$, $\sim 24$~K\cite{TrovarelliYbRh2Si2,CustersYbRh2Si2}, but much less than 200~K in $\beta$-YbAlB$_4$ which has more intermediate-valence trend\cite{MatsumotoYbAlB4}.
Since the lowest $T$ of the present study is significantly lower than $T^*$, Yb-124 certainly locates very close to the QCP.

 The QC of Yb-124 can be tested toward another axis, by looking at $H$ dependence of the properties.
In the panels of $M/H$ and $C/T$, 3$\sim$5~T along the easy axis $b$ works enough to disturb the non-FL state and changes it to the FL state below a few Kelvins.
The energy scale of $\mu_\text{B} H$ ($\mu_\text{B}$ is the Bohr magneton) is much less than $T^*$, and QC of Yb-124 is confirmed again at least within this scale.

Next, the properties of the Yb ion are discussed from the magnetization data  above $T^*$.
Clearly seen is strong Ising-type anisotropy along the $b$ axis in all the $T$ range [Fig.~\ref{fig:bulk}(b,f)].
The $M/H$ data for $b$ follow a Curie-Wiess law $M/H \propto (T-\Theta_\text{W})^{-1}$ well.
The effective moment $\mu_\text{eff}$ takes a large value of 5.5~$\mu_\text{B}$ per Yb
 when the fitting range is selected to 30~K$\sim$200~K [dotted line in Fig.~\ref{fig:bulk}(b)].
This $\mu_\text{eff}$ value considerably exceeds 4.54~$\mu_\text{B}$ which is the full value for the isotropic Yb$^{3+}$ ion (the total angular momentum $J=7/2$),
 and thus Yb-124 must have nearly pure $|J_z = \pm 7/2\rangle$ Ising GS doublet and the valence very close to +3
 (Note that $\mu_\text{eff}$ of an Ising spin is greater than that of an isotropic spin, as discussed in Supplemental Material with more quantitative details).
 This situation is very different from YbRh$_2$Si$_2$ at low $T$, having smaller $\mu_\text{eff}$ of 1.4~$\mu_\text{B}$ with the in-plane anisotropy\cite{GegenwartYbRh2Si2}. 
To our best knowledge, the effective moment of 5.5~$\mu_\text{B}$ for the GS doublet is the largest value in the QC materials,
 rather with a big difference. Of course, isotropic Ce$^{+3}$ cannot have $\mu_\text{eff}$ larger than 2.54.
More unusual is that large Ising spins lead to a high de Gennes factor and high transition $T$ of magnetism,
  and quantum fluctuations should be less effective due to small transition probabilities for whatever spin-related.
Exceptionally, the valence fluctuation could be unaffected by this problem.
 
 The Wilson ratio $R_\text{W} (=\pi^2k_\text{B}^2\chi/\mu_\text{eff}^2\gamma)$
  is found to be 2.7 from the  5-T data of $M/H$ and $C/T$ at the lowest $T$.
 The value of $\sim 3$ is slightly larger than 2, which is expected for ordinary heavy fermion metals with two-level GS degeneracy.
From this moderate value of $R_\text{W}$, negative $\Theta_\text{W}$, and no tendency of divergence of $M$ to the zero $T$ suggests that 
the QC fluctuations are related to the AF. This issue will be argued in the following microscopic study.  
\begin{figure}[htbp]
\centering
\includegraphics[width=0.9\linewidth]{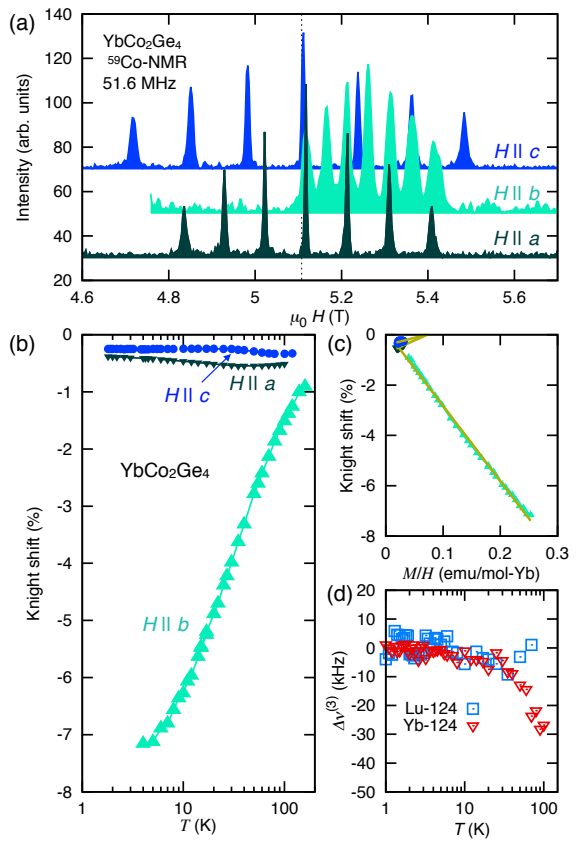}
\caption{\label{fig:nmr}(color online) {$^{59}$Co-NMR/NQR results for YbCo$_2$Ge$_4$.}% and LuCo$_2$Ge$_4$.
(a) Field-swept spectra at 40~K for $H \parallel a,b,c$.
Each spectrum consists of one central and six satellite peaks because the number of the equivalent site remains one even under the fields.
The dotted line indicates the Knight-shift origin for $^{59}$Co bare nuclei (${^{59}\gamma}H/2\pi$, where ${^{59}\gamma}/2\pi = 10.10213$~MHz/T).
(b) The Knight shifts ${^{59}K}$, which are static magnetizations at the Co nuclei, measured around at 5~T as a function of $T$.
${^{59}K}$ is resolved from the center-line spectral shift. 
 Higher-order perturbations from the quadrupole interaction are considered by the strict diagonalization.
(c) The Knight shifts plotted against $M/H$, to resolve the diagonal elements of the hyperfine coupling tensor.
The diagonal hyperfine elements ${^{59}A_{ii}^\text{hf}}$ are determined to be 0.57(5),-1.67(1),0.34(8) kOe$/\mu_\text{B}$, respectively.
(d) The $T$ dependence of the highest NQR line frequency $\Delta\nu^{(3)}(T) = \nu^{(3)}(T) - \nu^{(3)}$(0~K) for YbCo$_2$Ge$_4$ and LuCo$_2$Ge$_4$.
}
\end{figure}

\begin{figure}[htbp]
\centering
\includegraphics[width=0.9\linewidth]{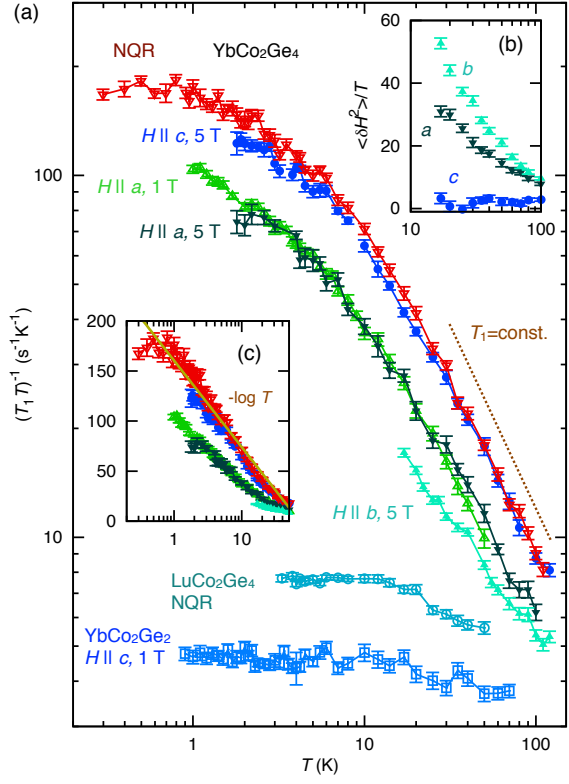}
\caption{\label{fig:t1}(color online) $^{59}$Co-NMR/NQR relaxation experiment 
(a) Relaxation rate $T_1^{-1}$ divided by $T$ for YbCo$_2$Ge$_4$, YbCo$_2$Ge$_2$, and LuCo$_2$Ge$_4$.
 The inset (b) shows the anisotropic local spin fluctuations at the Co nuclei resolved from the anisotropic $T_1^{-1}$ for YbCo$_2$Ge$_4$. 
 Relevance of small $\bm q$ fluctuations is inferred from this anisotropy plot (see the main text and the Supplementary Material).
 The inset (c) is a semi-log plot, indicating $-\log T$ behavior of $(T_1T)^{-1}$.
}
\end{figure}

Figure~\ref{fig:nmr} shows the single-crystal nuclear magnetic resonance (NMR) results for $^{59}$Co nuclei ($I=7/2$). 
The anisotropic Knight shifts ${^{59}K^{i}}$$(i=a,b,c)$, 
which is a measure of static magnetization at the wave vector $\bm q = 0$, reflects strong Ising anisotropy as expected [Fig~\ref{fig:nmr}(b)].
The measured ${^{59}K}(T)$ are generally composed of the $T$-independent (orbital, diamagnetism) chemical shifts, $K_\text{chem}$, 
and of the spin part $K_\text{s}(T)$:\\
${^{59}K^{i}}_\text{spin}(T)$ $= 
{{^{59}A_{ii}^\text{hf}}(\bm q=0)}M^i_\text{spin}(T)/{N_\text{A}\mu_\text{B}}H $,
where $N_\text{A}$ is the Avogadro's number, and ${^{59}A_{ij}^\text{hf}}$ is the hyperfine coupling tensor per Yb 1~$\mu_\text{B}$.
The $M^j_\text{spin}$ is the spin part of $M$. 
The ${^{59}K^b}-M^b/H$ curve crosses zero, and immediately $K_\text{chem} \ll K_\text{s}(T)$.
In addition, the very linear relation indicate that the magnetization of Co is not relevant in the total $M$
 and the ${^{59}}$Co-NMR measures mostly the $4f$ spins of Yb.

The electric field gradient (EFG)  at the Co nuclei was given by
 the spectral splitting of the NMR or the zero-field resonance, nuclear quadrupole resonance (NQR), frequencies (see Supplemental Material).  
At elevated $T$, in principle the EFG is changed by the phonon/CEF activations, lattice expansion, or valence fluctuation. 
In Fig.~\ref{fig:nmr}(d), $T$ dependence of the most significant NQR transition frequency ${^{59}\nu^{(3)}}$
 is plotted for Yb-124 and Lu-124.
 The ${^{59}\nu^{(3)}}$ of Yb-124 slightly decreases above $T^* \sim 23$~K.
We estimate the magnitude of the $\nu^{(3)}$ change by the Yb valence fluctuation by a point charge model.
The calculation found $\Delta \nu^{(3)}$ to be about +30~kHz when the Yb valence is increased by 0.03.
In terms of a valence transition or Kondo coherence of Yb systems, several NQR studies have been carried out.
For example, a few kHz of $\nu$ change was observed in YbCo$_2$Zn$_{20}$ as it enters to the FL state at 0.2~K\cite{MitoYbCo2Zn20}.
In Yb-124, the $\nu^{(3)}$ change occurs near $T^*$.
On the other hand, no difference is detected at lower $T$ within an order of 1~kHz, which corresponds to the 0.001 of Yb valence difference. Therefore the valence change itself seems not to have a QC feature.  

Following the valence fluctuation issue, 
the spin fluctuation of Yb-124 is focused on from the measurement of the NMR/NQR relaxation rate $T_1^{-1}$, 
which is shown in Fig.~\ref{fig:t1}(a).
The Korringa law, $ (T_1T)^{-1} =$const., applies for Yb-122 and Lu-124, because it is in FL state.
Yb-124 instead shows greatly enhanced spin fluctuations at low $T$, as a result of strong spin fluctuations or mass enhancement. 
The constant $T_1$ behavior above $T^*$ corresponds to a Curie law due to the localized Yb spins.  
 In more detail, ${^{59}T_1^{-1}}$ is proportional to local (at Co nuclei) magnetic fluctuations at measured frequency, $\langle (\delta H^\perp)^2(\omega) \rangle$.
The superscript $\perp$ means fluctuations perpendicular to the quantization axis,
 which is the external field direction for NMR or practically the $c$ axis for NQR of Yb-124, being effective to cause nuclear level transitions.
Immediately, a general expression like 
$\langle(\delta H^a)^2(\omega)\rangle \propto (T_1^{-1})^b + (T_1^{-1})^c - (T_1^{-1})^a$ is given.
Therefore, the anisotropic fluctuations can be resolved from $T_1$ measurement for three field directions,
as plotted in the inset (b) of Fig.\ref{fig:t1}.
The $T_1$ anisotropy experiment resulted in $\langle(\delta H^b)^2\rangle > \langle(\delta H^{a})^2 \rangle \gg \langle(\delta H^{c})^2\rangle$,
and presumably the small incommensurate AF wave vector would be the relevant spin fluctuations in the QC of Yb-124
(FM fluctuations along the easy axis cannot produce the magnetic fluctuations along the $a$ axis at the Co site in the given symmetry, see the Supplemental Material for the details).
The QC enhancement in $(T_1T)^{-1}$ is not prominent in the log-log plot, but $-\log T$ behavior is obvious in the semi-log plot [the inset (c)].

Up to now, the determined QC exponents are $T^{1.4}$ for $\rho_c$, and $-\log T$ for $\gamma = C_\text{M}/T$ and $(T_1T)^{-1}$.
These, except for $\rho_c$, are compatible with the class for two-dimensional AF\cite{SCRBook}, although the dimensionality, magnetic anisotropy,
 or a treatment for the incommensurate AF should be properly considered in further analyses.
In the series of the previously known Yb-based QC materials, peculiar correspondence has been discovered and stimulated theoretical efforts.
For example, YbRh$_2$Si$_2$, $\beta$-YbAlB$_4$, and the quasi-crystal Au$_{51}$Al$_{34}$Yb$_{15}$ exhibit diverging QC behavior in the uniform $\chi$, characterized by $T^{-0.6 \sim -0.5}$, $T^{1 \sim 1.5}$ in $\rho$, and $-\log T$ in $\gamma$\cite{CustersYbRh2Si2,MatsumotoYbAlB4,DeguchiYbQC}.  
All of these were successfully described by the valence fluctuation QC\cite{WatanabeVFQCPPRL}. 
However, Yb-124 has a distinct QC class, evidently because quantum fluctuations at the FM mode are absent.  
To summarize, we report new material YbCo$_2$Ge$_4$ and QC examined by the macro- and micro-scopic experiments.
The universality class of the QC would be two-dimensional AF class within conventional spin-fluctuation theories, but 
roles of the giant Ising-type GS of the Yb ions is still mysterious and has to be clarified. 
\begin{acknowledgments}
We thank Y.~Mezaki, M.~Takigawa, H.~Kato, N.~Katayama, Y.~Takano,
 and staff of the Materials Design and Characterization Laboratory in the ISSP for experimental support and discussions.
This work was supported by JSPS KAKENHI Grant No. 26707018, by Ito Foundation for the Promotion of Science,
 by  the Visiting Researcher's Program of the ISSP, and by MEXT Elements Strategy Initiative to Form Core Research Center.
\end{acknowledgments}
%\` %Just because of unusual number of tables stacked at end
\bibliography{document} % Produces the bibliography via BibTeX.

\clearpage
%\widetext
\begin{center}
{\bf\large Supplemental Materials}

\end{center}
\setcounter{equation}{0}
\setcounter{page}{1}
\makeatletter
\renewcommand{\theequation}{S\arabic{equation}}
%\makeatletter
\renewcommand*{\@biblabel}[1]{S#1.}
\makeatother
\setcounter{figure}{0}
\renewcommand{\thefigure}{S\arabic{figure}}%

\section*{Experimental details}
The growth conditions for $R$Co$_2$Ge$_4$ single crytals are described in the main text.
Excess flux was etched with hot $\sim$10\% hydrochloric acid.
The yield of 124 inside the ingot is typically not more than a few 10\%. 
What is more difficult is that ultrasonic cleaning or grinding degrades the crystals, probably because the 124 structure is metastable at room $T$.
Therefore we have never successfully acquired a powdered x-ray diffraction (XRD) pattern. 
These problems may have hindered identification of the 124 structure in past surveys:
Materials having exact 124 formula has not been reported so far in $R$-Co-Ge system.
However, Tm$_{0.15}$Co$_{0.30}$Ge$_{0.55}$ was previously observed during a
Tm-Co-Ge ternary phase diagram study as one of compounds with unknown structures\cite{FedynaTmCoGeTernary}, 
%isothermal sections at 870~K. 
 We guess that the reported Tm$_{0.15}$Co$_{0.30}$Ge$_{0.55}$ was identical to our 124 compound.
 
 Single-crystal XRD measurements were carried out on a CCD and a curved imaging plate
area detectors using a Mo-$K\alpha$ radiation.
No defect or inclusion of tin has been detected in the energy dispersive x-ray spectrometry (EDS) or single-crystal XRD analyses.
For $R=$Y and Tm, our information is limited to EDS and oriented XRD up to now, 
and only the $a$-axis lengths have been fixed to be 1.464 and 1.455~nm, respectively.

\begin{figure}%[htbp]
\centering
\includegraphics[width=0.8\linewidth]{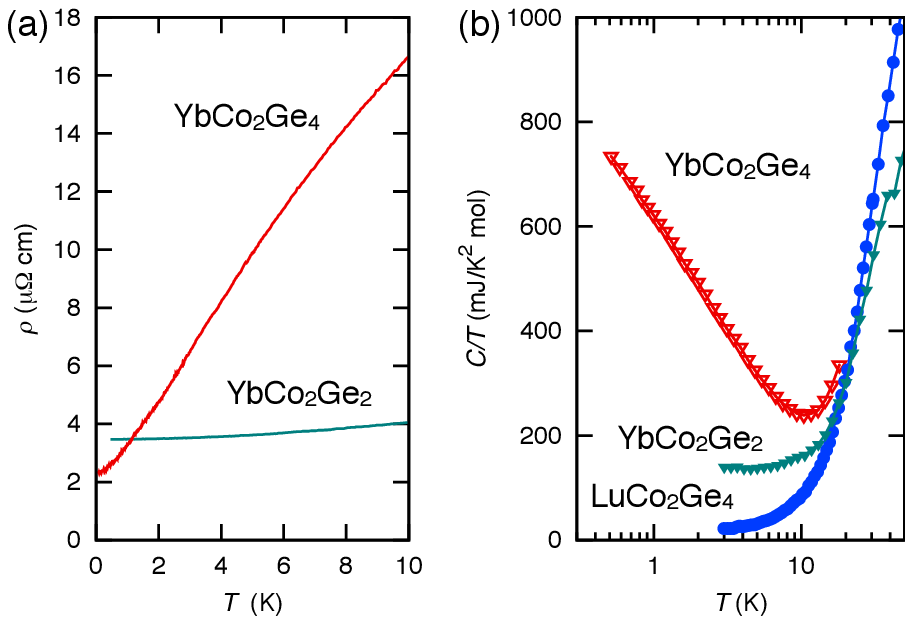}
\caption{\label{fig:bulk-si}(color online) (a) low $T$ expansion of the $\rho$ data in Fig.~{2(a)} of the main text.
(b) shows the $C/T$ of YbCo$_2$Ge$_4$, YbCo$_2$Ge$_2$, and LuCo$_2$Ge$_4$.
The $C_\text{M}$ in the main text is obtained by subtracting the data of LuCo$_2$Ge$_4$.
The present single-crystal result for YbCo$_2$Ge$_2$ is consistent with the previous polycrystalline experiment\cite{TrovarelliYbCo2Ge2}.
}
\end{figure}

The electrical resistivity $\rho$ was measured with a standard four-probe method in a dilution fridge, the magnetization was by 
a commercial SQUID magnetometer, and the specific heat $C$ was by a commercial magnet and ${^{3}}$He-fridge with a heat capacity option.
The in-plane $\rho$ in the low-$T$ range and raw $C$ before subtracting the phononic/$d$-electron part are shown in Fig.~\ref{fig:bulk-si}.
   
The NMR/NQR experiments have been performed with a standard pulsed spectrometer, 
and a cryo-cooled (liquid nitrogen) pre-amplifier to obtain better signal to noise ratio.
The NQR experiment below 0.9~K was carried out with a dilution fridge in Kyoto University. 
An transverse-field 7T NMR system with $^{3}$He/$^{4}$He circulation condensed by a 1.5-watt GM refrigerator was used in Kochi University for the other experiments. 
The NMR field-swept spectra were acquired by the fourier step sum method around at the center frequency.
The value of $T_1^{-1}$ was determined by 
fitting the time dependence of the spin-echo intensity of the central transition line after 
the inversion pulse to the theoretical formula\cite{NarathTiNMR}. For NQR with a finite $\eta (\sim 0.28)$, a recovery function is solved with the strict diagonalization: $\propto 0.206\exp(-3.0t/T_1)+0.566\exp(-9.302t/T_1)+0.228\exp(-18.37t/T_1)$.
For the easy axis $b$ below 15~K, line crossing caused by the large demagtization effect prevented relaxation measurent.  
Good fitting was obtained in the whole temperature range. 

\section*{On CEF level scheme of Yb ion and effective moment for Ising spin}
The $-\log T$ fit to the magnetic specific heat per Yb ion (see the main text) indicates that the entropy for the GS Kramers doublet is recovered above $T^*$.
Therefore, the higher CEF levels locate much above $T^*$.    
In addition, another evidence for absence of CEF levels between $T^*$ and $\sim$200~K is strong anisotropy observed in the magnetization measurement [see the main text, Fig.~{2}(b)]
 and Knight shift measurement [the main text, Fig.~{3(b)}].
The Curie-Weiss fit for the easy axis
\footnote{The Yb sites is on $4c$ Wyckoff position which has the two-fold rotational local symmetry along the $b$.
Then the local Ising axis cannot tilt from the $b$ axis,}, $M^b/H = N_\text{A}\mu_\text{eff}^2/3k_\text{B}(T-\Theta_\text{W})$ between 30~K and 200~K gives $\mu_\text{eff}$ of 5.5~$\mu_\text{B}$ and $\Theta_\text{W} = -20$~K.
This large moment should fully originate from Yb$^{3+}$, because Lu-124 is paramagnetic with small $\chi < 0.01$~emu/mol, and the $^{59}$Co NMR for Yb-124 does not detect the $\chi$ of Co\footnote{The on-site hyperfine coupling constant of the magnetic Co ion is typically an order of 100~kOe/$\mu_\text{B}$, but the experimental value of Yb-124 is 1.6~kOe/$\mu_\text{B}$, and the $K^b$ is almost perfectly proportional to the $M_\text{bulk}^b$.}.

A Curie-Weiss law for the two-level Zeeman splitting generally has the same form of the usual Curie-Weiss law for $J=1/2$
\footnote{Remind of a Curie law: $M =$\\
$ g_J \mu_\text{B}{\sum_{J_z} J_z \exp(g_J \mu_\text{B} J_z H/k_\text{B}T)}/{\sum_{J_z} \exp(g_J \mu_\text{B} J_z H/k_\text{B}T)}$\\
$\sim g_J^2 \mu_\text{B}^2 H {\sum_{J_z}  J_z^2}/{k_\text{B}T\sum_{J_z} 1 }  $}:
\begin{align}
M/H  \sim \frac{N_\text{A} g_J^2 \mu_\text{B}^2 J^2_\text{GS}}{k_\text{B}(T - \Theta_\text{W})},
\end{align}   
where $J_\text{GS}$ is the eigenvalue of $J_z$ for the GS$_+$ state and hereafter we take the quantization axis $z$ along the easy axis $b$.
If we adopt the generally-used above definition of $\mu_\text{eff}$, immediately 
\begin{equation}
\mu_\text{eff, Ising} = \sqrt{3}g_J \mu_\text{B} J_\text{GS},
\end{equation}   
Then the $\mu_\text{eff, Ising} = 6.93~\mu_\text{B}$ for $|\pm 7/2\rangle$ GS or $4.95~\mu_\text{B}$ for $|\pm 5/2\rangle$ GS
and these are indeed larger than the high-$T$ isotropic value of Yb$^{+3}$, $\mu_\text{eff}=4.54~\mu_\text{B}$.
From the experimental value of $5.5~\mu_\text{B}$ for 30~K$\sim$200~K, we obtain two limiting cases:
\begin{itemize}
  \item  $J_\text{GS} = 2.75$, and the valence of the Yb ion is 3+.
Since the diagonalized wave function $|\text{GS}_+\rangle$ does not have $|+ 5/2\rangle$ component, $\langle +7/2|\text{GS}_+\rangle > 0.79$.
  \item  $J_\text{GS}$ takes the maximum value of $7/2$, and the Yb ion is of intermediate-valence state.
The valence becomes $2 + 5.5/6.93 = 2.8$. 
\end{itemize}
% The anisotropy at the lowest $T$ ($>20$, from the Knight shift data which has much better sample alignment) puts another limitation.
%  The anisotropy can be estimated by 
%  \begin{equation}
% g_J|\langle GS_+| J_z | GS_+\rangle|/|\langle GS_+| J_{+} | GS_-\rangle|, 
%  \end{equation}
%  and then 
%  the mixing of $|\pm -5/2\rangle$ and $|\pm 7/2\rangle$ must be approximately smaller than 0.2
%  If $|\pm 5/2\rangle$ took the most significant part,  $J_\text{GS}$ cannot exceed 2.75.
Therefore, in any case, the major wave function of the GS should be $|\pm 7/2\rangle$. 

When a Curie-Weiss fit is applied instead to the higher $T$ range, between 200~K and 300~K, $\Theta_\text{W}$ takes a ferromagnetic value of 23~K, and $\mu_\text{eff}$ become decreased to $4.49~\mu_\text{B}$, 
which is close to the expected value without considering the CEF levels.
We infer that the next CEF energy levels are around at $\sim 200$~K.  
 \section*{NQR parameters}
 From the quadrupole splitting in the NMR spectra and the nuclear quadrupole resonance (NQR) observation at $\sim$2.88 and 4.46~MHz,
 the nuclear quadrupole coupling of Yb-124 were fixed.
 The traceless electric field (EFG) gradient tensor $V_{ij}$, which is proportional to the nuclear quadrupole coupling tensor, at low $T$ has a form of
\begin{align*}
\left|V_{ij}^{\text{exp}}\right|=
\begin{pmatrix}
  -1.33&  0&  0\\
  0&  -0.704&  \pm0.344\\
  0&  \pm0.344&  2.03
\end{pmatrix} \times 10^{21}\text{~V/m}^2.
\end{align*}
%  -0.965&  0&  0\\
  %0&  -0.51&  \pm0.25\\
  %0&  \pm0.25&  1.475 MHz
The $V_{ab}$ and $V_{ac}$ elements are zero due to the local symmetry. 
The largest eigenvalue of the NQR tensor, $\nu_Q$ = 1.505~MHz, and the asymmetric parameter $\eta$ = 0.28,
 and the principal axis for the $\nu_Q$ is in the $bc$ plane and 6.5\degree\ off axis from $c$.

A point charge model is inaccurate, but convenient method to obtain the EFG tensor by a calculation.
When the valences of Yb and Co are selected to be +3 and +0.86, and Sternheimer antishielding factor is -3,
the calculation within nearby 5 lattices can reproduce the above $\nu_Q$ and $\eta$, although the principal axes are different.
Such a discrepancy is usually based on itinerancy and/or covalency of conduction electron layers.
However, the roles of Yb ions in the EFG would be more like a point charge, 
and then we believe the effect of the valence change could be more accurate in the point charge calculation.

 \section*{On anisotropic spin fluctuations at the Co sites}
% and is reproduced by NQR although the $T$ range quite narrow, 2~K $< T<$ 10~K.
% The Korringa behavior again applies when $T < 1$~K. However, 
% the crossover $T$ to the FL state derived by the above bulk experiments is $<$0.2~K, and this discrepancy is an open issue.      

$^{59}$Co NMR/NQR is a local probe, and $T_1^{-1}$ provides solely $\bm q$ averaged information if Co carried all the magnetic moments.
Nevertheless, the local symmetry of Co $8e$ site and relations to the adjacent Yb moments, aka a form factor, allows us to unravel the fluctuating spin structures. 
When the Yb moments are collinear along the external field, 
the ${^{59}A_{ij}^\text{hf}}$ tensor for $\bm q = 0$ in the 124 lattice become a form of 
\begin{align}
{^{59}A_{ij}^\text{hf}}(\bm q=0) = 
\begin{pmatrix}
A_{aa}& 0& 0\\
0& A_{bb}& A_{bc}\\
0& A_{bc}& A_{cc}
\end{pmatrix}.
\end{align}
The $A_{ab}$ and $A_{ac}$ elements are zero  (Co has a two-fold rotation symmetry parallel to the $a$ axis), 
and therefore the local fluctuations along the $a$ axis cannot be produced even when the Yb moments fluctuate along the $b$ (or $c$) axis.
%And considering $M/H$ displays large anisotropy to the $b$ axis and the local symmetry at the Yb $4c$ site, 
%the local Ising axis of Yb has to be exactly the same direction of the magnetization easy axis, $b$.
Eventually, $\langle(\delta H^{b,c})^2\rangle \gg \langle(\delta H^a)^2\rangle$ for the FM mode, $\bm q = 0$. 
This relation apparently contradicts to the results in the main text of Fig.~{4}.
Therefore, finite $q$ fluctuations have to be considered.

For the commensurate AF fluctuations, and for simplicity, an obvious case [$\bm q=(0,0,1)$] is easily formulated within the nearest neighbors: 
\begin{align}
{^{59}A_{ij}^\text{hf}}\left(0,0,1\right) = 
\begin{pmatrix}
0& A_{ab}& A_{ac}\\
A_{ab}& 0& 0\\
A_{ac}& 0& 0
\end{pmatrix},
\end{align}
although an estimation of the values for $A_{ab,ac}$ would be impossible.
A similar way was applied in the $^{75}$As-NMR experiment of
 the iron-based superconductivities where commensurate AF fluctuations are relevant\cite{KitagawaBa122}.
In this case [$\bm q=(0,0,1)$], solely the off-diagonal elements are important, contrary to the FM case.
Then, $\langle(\delta H^a)^2\rangle \gg \langle(\delta H^{b,c})^2\rangle$ for $\bm q=(0,0,1)$ fluctuations.

Experimentally,
$\langle(\delta H^a)^2\rangle : \langle(\delta H^{b})^2 \rangle : \langle(\delta H^{c})^2\rangle \sim 3:5:0$.
Considering the local Ising axis of the Yb ions is $b$, 
\begin{align}
{^{59}A_{ij}^\text{hf}}\left(\text{the relevant }\bm q\right) \propto 
\begin{pmatrix}
\text{n/a}& 3& \text{n/a}\\
3& 5& 0\\
\text{n/a}& 0& \text{n/a}
\end{pmatrix}.
\end{align}
Therefore, the underlying fluctuation mode has to be characterized between the above two cases. 
A plausible candidate is one with incommensurate $\bm q$ close to 0, as YbRh$_2$Si$_2$ has $\bm q = (0.14,0.14)$ fluctuations
 near the field driven QCP\cite{StockINSYbRh2Si2}. 
 For YbRh$_2$Si$_2$, $^{29}$Si-NMR study also suggested a presence of AF fluctuations since the $(T_1T)^{-1}$ obeys $T^{-0.5}$ law to the lowest $T$ while $K$ is constant\cite{IshidaYbRh2Si2}.
It is difficult to tell the exact wave vector for such an incommensurate AF mode by the present NMR study,
 but it would be possible if some of the off-diagonal component in ${^{59}A_{ij}^\text{hf}}$ are solved by low-field NMR experiments in future studies. 

Finally we note that the relevance of the incommensurate AF fluctuations with the wave vector close to 0 is reasonably consistent with the moderately large Wilson ratio of 2.7.  

\end{document}